# Quantum-Chiplet: A Novel Python-Based Efficient and Scalable Design Methodology for Quantum Circuit Verification and Implementation

Yu-Ting Kao [1], Hao-Yu Lu [2], Yeong-Jar Chang[1], Darsen Lu[2,3]

[1] Industrial Technology Research Institute, [2] Intelligent Computing Industrial Doctorate Program, Miin Wu School of Computing, National Cheng Kung University, [3]Institute of Microelectronics, Department of Electrical Engineering, National Cheng Kung University

Correspondence author: SunnyKao@itri.org.tw

*Abstract*—Analysis and verification of quantum circuits are highly challenging, given the exponential dependence of the number of states on the number of qubits. For analytical derivation, we propose a new quantum polynomial representation (QPR) to facilitate the analysis of massively parallel quantum computation and detect subtle errors. For the verification of quantum circuits, we introduce Quantum-Chiplet, a hierarchical quantum behavior modeling methodology that facilitates rapid integration and simulation. Each chiplet is systematically transformed into quantum gates. For circuits involving n qubits and k quantum gates, the design complexity is reduced from "greater than $O(2^n)$" to $O(k)$. This approach provides an open-source solution, enabling a highly customized solution for quantum circuit simulation within the native Python environment, thereby reducing reliance on traditional simulation packages. A quantum amplitude estimation example demonstrates that this method significantly improves the design process, with more than 10x speed-up compared to IBM Qiskit at 14 qubits.

## I. INTRODUCTION

Quantum computing offers significant advantages due to superposition and entanglement [1-4]. However, despite these advantages, the simulation and design of large quantum circuits remain a significant challenge, as the complexity grows exponentially with the number of qubits. For example, a 5-qubit system requires managing 32×32 matrices and 32×1 vectors [6-7], which dramatically increases both design time and engineering resources. Therefore, there is a need to explore new methods to significantly reduce this complexity.

In this paper, we address this gap with a novel approach to implement quantum circuits with vector and matrix operations in Python. We demonstrate that the NumPy package for matrix and vector computations effectively supports noise-free simulations, replacing the functionality provided by traditional quantum simulation packages, such as IBM Qiskit [10]. We further show that large quantum circuits, such as quantum amplitude estimation (QAE), can be easily designed and simulated within the native Python environment with a flattened gate-level design complexity of $O(k)$.

In this paper, we first introduce in section II-A the Quantum Polynomial Representation (QPR), which is a analytical approach for quantum circuit derivation, and also an enhancement to the traditional Dirac notation, since both QPR and traditional Dirac notation have exponentially increasing vectors and matrices. QPR offers an intuitive visualization of quantum states, making it easier to identify hard-to-detect errors.

In section II-B, for quantum circuit implementation, we propose a novel method, Quantum-Chiplet, which not only simplifies the design process, but also leverages the conventional circuit design concept of behavior-level modeling in the early stages – a feature not supported by existing quantum simulation packages, allowing the design complexity to be improved to $O(k)$ without using Qiskit. The advantage of behavioral-level modeling is that it does not require detailed quantum gates and can quickly integrate large circuits. Because the circuits are too complex to use Qiskit, the design complexity is greatly improved. Once the behavior-level model is verified, the design progresses to subsequent stages of quantum gate implementation.

By utilizing QPR and Quantum-Chiplet, we streamline the quantum design process and enhance the efficiency of quantum algorithm implementation. The effectiveness of this approach is demonstrated through QAE, with quantum advantages [8], in section II-C. This methodology contributes to efficient and accurate circuit design, enabling simulations with a fully open source solution for quantum circuit design. The implementation details are shown in section II-D. Finally, the advantages in terms of computational speed compared to IBM Qiskit is quantified in "Section III - Results".

## II. METHODOLOGY

### A. Quantum Polynomial Representation (QPR)

For today's digital VLSI circuits, when an input signal set $v_1$ results in an output $o_1$, and a different input signal set $v_2$ results in an output $o_2$, there is no superposition relationship between the two. However, since quantum circuits are linear systems, superposition principle applies. For example, when the input is the sum of the two vectors, $v_1 + v_2$, the result would become $o_1 + o_2$. We take advantage of this by applying a large number of superimposed signals, thereby creating massive parallelism in computation. QPR proposed in this paper aims to assist circuit designers by providing a clear framework to analytically derive the outcome of quantum computation with superposition in consideration. QPR not only supports quantum Pauli gates for bit flipping, but also supports quantum operations for multiple qubits, such as the Hadamard gate.

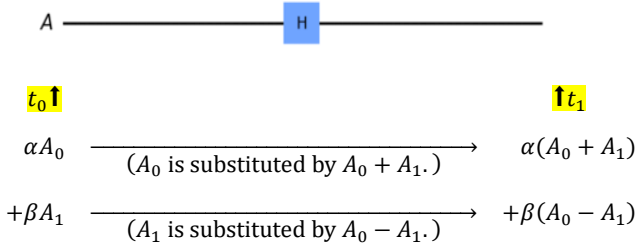

Fig.1. QPR variable substitutions for Hadamard gate

QPR is obtained by inserting qubit names explicitly into quantum state representation. For example, with qubit A, the QPR for Dirac notation $|0\rangle$ is $A_0$, and that of $|1\rangle$ is $A_1$. Operations of quantum gates correspond to variable substitutions in QPR. For example, the Hadamard gate operates on both the "0" components ($A_0$) and the "1" component ($A_1$) of qubit A, respectively, and adds the outcomes together following superposition principle, as illustrated in **Fig. 1**. The advantage of QPR is that it clearly expresses the state of each qubit, avoiding confusion during analytical derivation. Notice that we have neglected the normalization factor of $1/\sqrt{2}$ for simplicity. Normalization can be done later whenever needed. **Fig. 2** shows another example of QPR-based analysis of the "A circuit" part of QAE, consisting of two Hadamard gates and one controlled-NOT (CX) gate. With QPR notation, all the possible states within the computation can be visualized clearly, by means of polynomial expansion, i.e., the expansion of product of sums into sum of products. Polynomial expansion is not easy to do with conventional Dirac notation. The full circuit (**Fig. 4**) will be discussed in Section II-C.

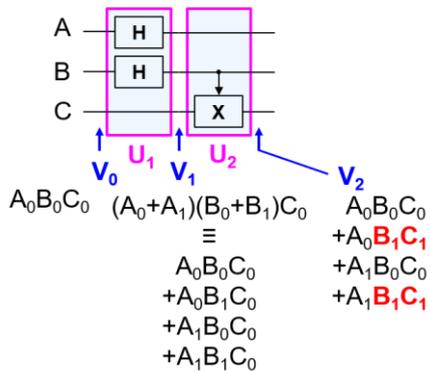

Fig. 2. The main operation (A circuit) of QAE is analyzed by deriving intermediate states ($V_0$, $V_1$, and $V_2$) using the QPR method.

To give a third example, we analyze, with QPR, a quantum circuit that performs analog multiplication and analog addition, as pre-operations of QAE [5]. While the multiplication operation was correctly performed, the addition operation was erroneously derived, as we will show later. Derivation of the circuit outcome is quite complex with conventional Dirac notation. The QPR method proposed in this paper greatly simplifies the derivation, and the errors can be identified easily. **Fig. 3** shows the multiplication and addition circuit [5]. The original circuit (**Fig. 3(a)**) is modified to correct addition operation, resulting in the circuit in **Fig. 3(b)**. The original Dirac notation does not clearly display the states of B and C, which makes it difficult to identify these errors. In contrast, these errors can be more easily detected using QPR notation.

With QPR, we can find out that h(x) is activated when C=1, g(x) is activated when C=0, and the state of L changes only when B=1 and C=0. Therefore, the initial state of $B_0C_0L_0$ becomes $B_0(C_0 + C_1)L_0$ after passing through the first Hadamard gate. The "h" gate operates on qubit B when C=1, so that $B_0C_1L_0$ becomes:
$$(B_0\sqrt{1 - h(x_i)} + B_1\sqrt{h(x_i)})C_1L_0,$$
$i$ represents the index, which is used to distinguish different inputs or calculation steps. For example, if there are multiple input data points $x_1, x_2, x_3, \ldots$, then $i$ is used to label these different data points. $x_i$ represents the $i$th input variable, i.e., the input value corresponding to index $i$, which affects the behavior of quantum gates $g(x_i)$ and $h(x_i)$.

whereas "g" gate operates on qubit B when C=0, so that $B_0C_0L_0$ becomes:
$$(B_0\sqrt{1 - g(x_i)} + B_1\sqrt{g(x_i)})C_0L_0.$$
After passing through the second Hadamard gate, the full expression becomes
$$\sum\left\{\left[B_0\sqrt{1 - h(x_i)} + B_1\sqrt{h(x_i)}\right]((C_0 - C_1)L_0) \right.$$
$$\left. + \left[B_0\sqrt{1 - g(x_i)} + B_1\sqrt{g(x_i)}\right]((C_0 + C_1)L_0)\right\}$$
Subsequently, the CCX gate operates on only the $B_1C_0L_0$ state to generate the following term:
$$\sum[\sqrt{g(x_i)} + \sqrt{h(x_i)}]B_1C_0L_1$$
As the sum of the probability of occurrence of all terms is the total probability, i.e., $p = \sum(p_i) = \sum(c_i^2)$ where $c_i$ is the coefficient of each term. As the term above is the only one with L=1, the total probability of measuring L=1 becomes,
$$p(L = 1) = \sum\left(\sqrt{g(x_i)} + \sqrt{h(x_i)}\right)^2$$
$$= \sum\left(g(x_i) + h(x_i) + 2\sqrt{g(x_i)h(x_i)}\right)$$
Since the $2\sqrt{g(x_i)h(x_i)}$ terms cannot be eliminated, the measured probability is not the same as probability addition $g(x_i) + h(x_i)$.

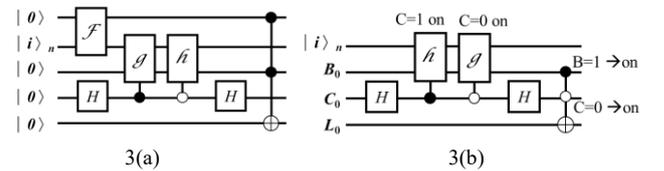

Fig. 3. Multiplication and addition circuit [5]: (a) original circuit (b) modified circuit for correcting the addition part

### B. Quantum-chiplet – Simulating and Designing Quantum Circuits with Matrices

To simplify the design process of large-scale quantum circuits, this paper proposes a behavioral level modeling approach, where we use matrix multiplications to represent a combination of quantum gates operation. The application, or problem in question, is defined with a unitary operation matrix (U). The matrix can subsequently be converted into quantum gates using a Design Compiler. The Design Compiler then facilitates the completion of the digital design, converting the matrix into digital quantum gates. The implemented results will be presented in a separate publication. We have named such design approach as "Quantum-Chiplet," which refers to the fact that quantum gates (e.g., X, CX, CCX, H, Z, R) are stacked modularly like building blocks.

We define the behavior model as a matrix operation to achieve the following key advantages. First, multiple matrices can be pre-calculated and merged into a single matrix. Second, complex operations of many quantum gates can also be represented as a single matrix. As a result, the complex quantum system can be efficiently implemented and simulated using these behavior models. This approach enables clear and efficient quantum circuit logic design.

For traditional quantum circuit simulators, the matrix representation of quantum gates are fixed and cannot be modified. On the other hand, in our proposed methodology, we directly use matrices for simulation, and its content can be modified arbitrarily during the design process. This shortens the time needed for building up large quantum circuits with complicated functions.

After design and verification, the behavior model is translated into quantum gates by breaking the matrix into specific quantum operations, constructing a "real" quantum circuit with all required gates. This is similar to what a design compiler does in conventional digital VLSI circuit design.

One of the primary advantages of the Quantum-Chiplet method is that it is an open-source framework, which provides researchers and developers with unrestricted access to its architecture, facilitating community-driven innovation, customization, and optimization for diverse computational requirements.

Moreover, the Quantum-Chiplet method exhibits notable flexibility and scalability, making it suitable for a broad spectrum of applications. By eliminating dependence on proprietary commercial platforms, this approach lowers the entry barriers to quantum computing research and accelerates the adoption of quantum technologies. As the Quantum-Chiplet framework continues to be refined and validated, it holds the potential to become a foundational tool in the open-source quantum computing ecosystem, offering an efficient and scalable alternative for academic and industrial research. We demonstrate the effectiveness of Quantum-Chiplet with a large QAE circuit [8, Fig. 1] in the next section.

### C. Quantum Amplitude Estimation (QAE) Circuit Design

The QAE system primarily consists of three components (**Fig. 4**). The first part is the A circuit, which serves as the main operational block. The Control_Q circuit manages measurement accuracy by exploiting the inherent periodicity of quantum states to improve the resolution of amplitude detection, while the inverse QFT converts phase information into frequency components, allowing us to deduce the amplitude value of a specific quantum state through measurement. QAE performs m repetitions of operation A to obtain the average of $2^n$ computations for n qubits. Compared to traditional Monte Carlo methods requiring $m^2$ iterations for the same accuracy $1/m$ [12], QAE offers a quadratic advantage through quantum technology.

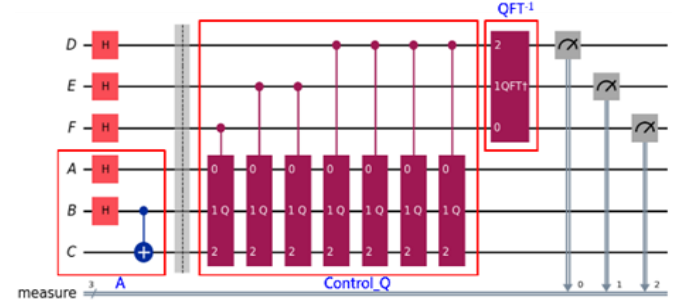

Fig. 4. Implementation of QAE circuit in Qiskit, with three major components highlighted.

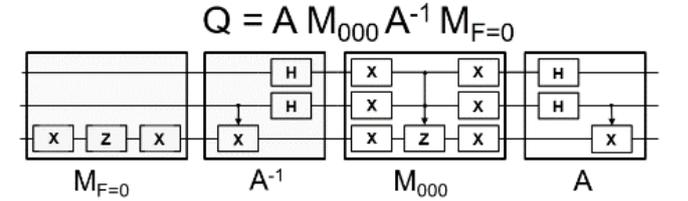

Fig. 5. Control_Q circuit composition

- Control_Q circuit

First, we define a single Q circuit without a control point. (**Fig. 5**).
$$Q = (A\ M_{000}\ A^{-1})M_{F=0}$$
Then, we target $Q^1$, $Q^2$, and $Q^4$, and add a control point to each Q circuit, as shown in Figure 4. Each control_Q contains a single Q circuit and an external control point. It is activated when Control_point = 1 and remains inactive when Control_point = 0. These qubits are sent to the inverse QFT module. The QFT utilizes frequency-based methods to optimize the efficiency and precision of amplitude estimation. By leveraging the periodicity inherent in quantum states, the QFT enhances the resolution of amplitude detection, which is crucial for accurately extracting amplitude information from the quantum system.

- Inverse Quantum Fourier Transform (QFT$^{-1}$)

The primary purpose of the inverse QFT is to convert quantum phase information into a binary representation, enabling it to be read through measurement efficiently. Additionally, QFT translates phase information into frequency components,

$$QFT^{-1}: \{A_n\} \mapsto \frac{1}{\sqrt{N}} \sum_n A_n e^{-jwn} \quad (1)$$
$$FFT: \{a_n\} \mapsto \sum_n a_n e^{-jwn} \quad (2)$$

It allows us to calculate the amplitude of the quantum state via measurement, thereby ensuring more accurate amplitude estimation. In matrix operations, the Fast Fourier Transform (FFT) can be employed as a substitute for inverse QFT to achieve the same effect.

## D. Quantum-chiplet Implementation

After realizing all quantum gates, both QPR and Dirac notation require the design of $2^n \times 2^n$ matrices and $2^n \times 1$ vectors for simulating n-qubit circuits. This results in an exponential increase in design complexity, which impedes engineers from completing the circuit design. The proposed method, Quantum-Chiplet, involves treating the basic unit quantum gates as a $2 \times 2$ matrix, which is analogous to a single Quantum-Chiplet. If the quantum circuit is constructed by stacking these Quantum-Chiplets, the design complexity can be reduced from $O(2^n)$ to $O(k)$ (n qubits, k gates). Notice that the simulation complexity is still unchanged ($O(2^n)$), despite that the design process, which requires human intervention, is greatly simplified.

Taking **Fig. 6** as an example, the quantum gates are stacked together using the Kronecker product ("kron" operation in the Python *NumPy* library). Each operation results in a new matrix with doubled dimensions. Matrix multiplication is implemented with the python "dot" operation. In **Fig. 6**, vectors $V_0$, $V_1$ and $V_2$ represent the quantum state at the indicated positions, matrices $U_1$ and $U_2$ represent the operation matrix of the quantum gates, the $I$ matrix is a $2 \times 2$ unit matrix for dimensionality enhancement, and the $H$ matrix is a $2 \times 2$ matrix for a single Hadamard gate. This circuit is composed of 3 qubits, so each $U$ operation matrix is an $8 \times 8$ matrix. $V_0$ represents the initial state of this circuit, $A_0 B_0 C_0$. It is also represented by the vector $[1,0,0,0,0,0,0,0]^T$. $U_1$ is an $8 \times 8$ operation matrix for $H_A \otimes H_B \otimes I_C$. $U_2$ is an $8 \times 8$ matrix for the CX gate, which is obtained by the kron operation of $I$ and the original 4x4 CX gate matrix. The CX gate allows a specified qubit state to control CX. Here, CX inverts the "C" state only if B=1. Notice that in Quantum-Chiplet the "kron" operation ensures consistent operation dimensions.

Quantum-Chiplet pre-calculates the operation of multiple matrices and saved the result as one matrix to speed up the overall design procedure. The saved matrix can replace the original multiple matrices, resulting in accelerated simulation as n becomes large, and in the presence of repeatedly operations. It also allows a simple hierarchical design procedure, as is typically done for digital VLSI design. For example, in the implementation of QAE as described in the previous section, the matrix representing the control-Q circuit is first calculated, and then repeated 7 times, with minor modification to the control signal.

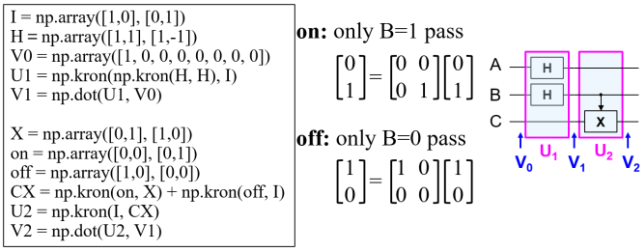

Fig. 6. Quantum-Chiplet using the kron-and-dot method

## III. RESULT

Taking the QAE circuit as an example, comparing the execution time of *Qiskit* and *Quantum-Chiplet*, it can be seen that when the number of qubits is more than 10, the execution time of *Qiskit* is much longer than that of Quantum-Chiplet (**Fig. 7**).

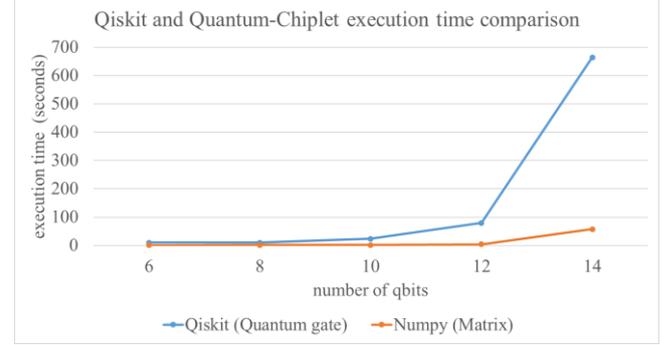

Fig. 7. Qiskit and Quantum-chiplet (NumPy) execution time comparison

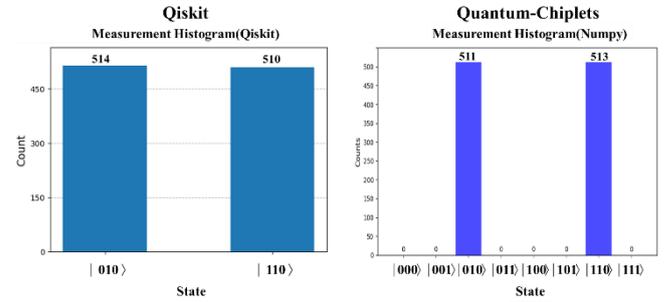

Fig. 8. Histogram Analysis of Qiskit and Quantum-Chiplet

This study uses matrix-based design and the Quantum-Chiplet methodology, which not only achieves the same result as Qiskit, but also significantly enhances simulation efficiency. Taking QAE as an example, at 14 qubits, for example, the Qiskit execution time is 664.34 seconds, while Quantum-Chiplet design executes in 57.24 seconds. This demonstrates a simulation efficiency enhancement of 10 times or larger.

Histogram analysis is a common feature of quantum circuit simulation. **Fig. 8** shows that the histogram analysis results for both methods are nearly identical, reflecting the same underlying probability distribution. From this measurement, the average value after $2^n$ parallel computations can be inferred, eliminating the need for multiple measurements as in histogram analysis.

## IV. CONCLUSION

This paper proposes QPR, utilizing polynomial representation to facilitate simple and efficient analytical derivation for quantum circuits. Additionally, a matrix-based behavior level design methodology is introduced, offering a streamlined approach to quantum circuit design. In the early stage of design, high-level matrix-based behavior models were used to integrate different components into large circuits. They are compiled into actual quantum gates at the later stage of design. When all design blocks have become quantum gates, designs and simulations can be completed in the open-source Python environment with the novel

Quantum-Chiplet methodology. The design complexity is reduced from $O(2^n)$ to $O(k)$ with n qubits and k gates, eliminating the need and reliance on traditional quantum circuit simulators. More than 10-fold simulation time reduction as compared to the *Qiskit* package has been demonstrated. This approach provides a self-contained, efficient, and scalable solution for quantum circuit design and simulation.

## REFERENCES

[1] C. H. Bennett, G. Brassard, "Quantum cryptography: Public key distribution and coin tossing," Proceedings of IEEE International Conference on Computers, Systems and Signal Processing, Bangalore, India, 1984, pp. 175–179.

[2] A. Y. Kitaev, A. H. Shen, M. N. Vyalyi, Classical and Quantum Computation, American Mathematical Society, 2002.

[3] L. K. Grover, "A fast quantum mechanical algorithm for database search," Proceedings of the 28th Annual ACM Symposium on Theory of Computing (STOC), 1996, pp. 212–219.

[4] M. Nielsen, I. Chuang, Quantum Computation and Quantum Information, Cambridge University Press, 2010.

[5] A. Carrera Vazquez and S. Woerner, "Efficient State Preparation for Quantum Amplitude Estimation," Physical Review Applied 15 (2021), 10.1103 / physrevapplied.15.034027.

[6] D. DiVincenzo, "Two-bit gates are universal for quantum computation," Physical Review A, vol. 51, no. 2, 1995, pp. 1015–1022.

[7] A.Poornima, N. N. M., and R. Ujjinimatad, "Matrix representations for quantum gates," International Journal of Computer Applications, vol.159, no.8, Feb. 2017.

[8] Nikitas Stamatopoulos et al., "Option Pricing using Quantum Computers", Quantum, 4, 291, Jul. 2020.

[9] Lu, C., Pilato, C., & Basu, K. (2023, April). Towards High-Level Synthesis of Quantum Circuits. In 2023 Design, Automation & Test in Europe Conference & Exhibition (DATE) (pp. 1-6). IEEE.

[10] Qiskit Development Team. (n.d.). Qiskit documentation. Qiskit.org. Retrieved December 9, 2024, from https://qiskit.org/documentation/

[11] Smith, K. N., Ravi, G. S., Baker, J. M., & Chong, F. T. (2022, October). Scaling superconducting quantum computers with chiplet architectures. In 2022 55th IEEE/ACM International Symposium on Microarchitecture (MICRO) (pp. 1092-1109). IEEE.

[12] Reuven Y. Rubinstein, *Simulation and the Monte Carlo Method,* Wiley Series in Probability and Statistics (Wiley,1981).